# IMPORTANCE OF THEORY, COMPUTATION AND PREDICTIVE MODELING IN THE US MAGNETIC FUSION ENERGY STRATEGIC PLAN

Fatima Ebrahimi (PPPL/PU), Gary Staebler (GA), Paul Bonoli (MIT), Francois Waelbroeck (UT), Chris Hegna (UW-Madison), Lynda LoDestro (LLNL), M.J. Pueschel (UT), I. Joseph (LLNL): Based on the community input at the NAS Madison and Austin workshops (July and December 2017)

## 1.   Goals and Description of the Initiative:

Due to the complexity of fusion plasma, materials, and nuclear science, remaining gaps in our scientific understanding of the underlying physics represent some of the most fundamental challenges for achieving a viable fusion reactor. Models of self-heated burning plasmas confined by magnetic fields require understanding the nonlinear interaction of multiple physics processes over spatial and temporal scales spanning many orders of magnitude. To overcome these multi-physics and multi-scale challenges, developing fundamental understanding through theory and computation, combined with advances expected in high performance and extreme-scale computing in the coming decade, is necessary for accelerating the development of a fusion-based energy source. At the two recent NAS community workshop series (in July and December 2017) "Theory and Computation" was recognized as a key strategic element that constitutes a natural foundation for any US strategic approach toward the development of magnetic fusion energy. In the US MFE program, the five main missions of this element are to: **1- continually broaden and deepen our understanding of the physics of fusion burning plasmas, 2- develop physics-based validated predictive capability, 3- discover new modes of operation, 4- explore and optimize device design, and 5- develop real-time plasma control systems**.

To fulfill these objectives, the following components are essential tools for the MFE theory program:
- Analytical theory, supported by semi-analytical techniques and simplified models
- A multi-fidelity hierarchy of physics models including
  - large-scale high-fidelity simulations for applications such as whole-device integrated modeling,
  - reduced integrated models calibrated by highest fidelity physics simulations, and experimentally validated for fast prediction and real-time control,
  - standalone computational studies to generate both fundamental knowledge (which can then inform the other two components above) and verification/validation.

The enabling capabilities through innovative analytical techniques, reduced models and advances in high performance computing could lead to fundamental understanding, discovery, and real-time control as well as
- closing some of the remaining gaps for reliable prediction for burning plasmas including ITER
- optimization of tokamak, stellarator and other alternative MFE concepts, in order to achieve the mission of achieving fusion energy within the next few decades
- the ability to quickly identify and efficiently counteract potential unforeseen roadblocks that may arise for any given path to fusion energy.

Historically, the fundamental scientific impact of fusion theory has extended far beyond the MFE community. There are many examples of conceptual advances, pioneering analytical techniques and high-end computational plasma physics models developed in the US MFE community that have also contributed to significant advances in other subfields of plasma physics as well as in the international MFE programs.

2. **Programmatic Benefits:**

Some major areas of any MFE strategic plan that benefit from theory and computation are:
- **Deployment of predictive theory as a tool for discovery and support of existing experiments**: The US has invested in a world leading diagnostic measurement capability that has greatly advanced the validation of fusion plasma theory leading to the development of predictive reduced models. This fusion energy science mission has strengthened the confidence in the success of ITER and is now being utilized by China to design their next step CFETR machine. This science mission is essential to a successful US strategic plan to develop fusion energy within the next few decades. Existing MFE tokamaks, DIII-D and NSTX-U, are highly diagnosed and provide detailed multi-scale validation data for physics models. Physics codes with synthetic diagnostics enable far more detailed analyses and interpretation of experimental results. Efficient validation and analysis workflows have now made the uncertainty quantification of large data sets possible. The resulting confidence in theoretical predictions will enable new discoveries ranging from innovative control methods to the unforeseen modes and regimes of operation that they will enable.
- **Next large-scale US fusion experiments**: The US fusion energy science program has developed validated predictive models that will enable a better-informed selection of the next generation of US fusion experiments. The predictions made through high performance computing and modeling can be used to identify specific configurations that are designed to meet performance goals, yet stay within operational performance limits, for example confinement time or pulse length, and optimize the ultimate viability of fusion reactors. Theory and predictive computing, whether used for the performance extension and control scenarios on existing large-scale devices or for design innovations to be explored on next step experiments, will reduce the cost and shorten the timeline of the US path forward for fusion energy and provide guidance for a strategic plan.
- **ITER**: With ITER now under construction, the next decade provides many new opportunities for theory, computation and predictive modeling to have an impact on critical aspects of burning plasma operation, including tolerance of heat and force loads on the first wall, and control of transient events. Prediction of ITER operation from discharge startup to ramp-down is a strategic goal for the US program that will enable improving the fusion performance of ITER. By refining theoretical models based on ITER performance, and by confirming the predictive capability of existing models, confidence will be increased that the US possesses the tools of designing next-step devices beyond ITER using a solid physics basis.
- **Cross-field interaction and educational enrichment**: Theory and computation have had a significant role in promoting synergy between the fusion program and other branches of plasma physics research. Pioneering theories such as spontaneous tearing reconnection or kinetic sub-scale gyro-averaged models, which are applicable in many branches of plasmas physics, have been first initiated and established within the fusion community. Advances in computation and theory will further promote cross-field interaction in plasma physics as a field that unifies fusion, low temperature, high-energy-density, space, and astrophysical sciences. This interaction enriches both the educational and scientific aspects, and attracts younger students and scientists to pursue their careers in developing the groundbreaking solutions necessary to achieve the mission of developing a fusion power reactor.
- **Leadership in fusion and plasma science:** Theory and computation has greatly contributed to the US leadership in fusion and plasma science. Some recent *scientific advances in fusion* enabled through combined theoretical and experimental effort, and strong and essential partnership between the USA and Europe are:
  - *theoretical prediction and experimental demonstration of neoclassical tearing mode stabilization by localized electron cyclotron current drive [1]*
  - *understanding and quantitative verification of global mode stability in experimental high performance tokamak plasmas, based on drift-kinetic MHD theory [2]*

A defining characteristic of the US fusion program has been its strong emphasis on continually advancing the *frontiers of plasma physics*. Two general areas of US leadership enabled through strong engagement of fusion theorists are:
- *leading-edge plasma-physics research through NSF-DOE partnership*
- *high-end computing at NERSC and the Leadership-class facilities and the establishment of fusion integrated simulation through SciDAC/ASCR partnership*

The advent of computation at the exascale in the US presents opportunities to advance all areas of plasma and fusion material science. To maintain a leading role for the US in the pursuit of a viable controlled thermonuclear reactor, it is critical to maintain, support and encourage the interplay between reactor design and high-quality, *leading-edge plasma-physics research and computer science*. Major efforts must be placed on developing codes and capabilities for simulating plasma behavior. These efforts place great demands on computational methods, advanced algorithms and hardware, and are undoubtedly useful well beyond fusion physics.

## 3. **Current Status and US Leadership:**

High performance computing (HPC) is critically important for our present US fusion theory program. Over the past decades, the US fusion program has led the world in developing new physics models in the areas of gyrofluids, gyrokinetics, wave-particle interactions and extended MHD, and numerical methods to exploit advanced computing. In particular, numerous SciDAC/ASCR partnerships over the previous decade facilitated advances in high-performance computing using petaflop architectures. To maintain this scientific leadership position additional pioneering plasma science is critical in the areas of 1) analytical theory for the understanding of the physics of plasmas, for the development of the basis of computational models, as well as to interpret, reinforce and verify computation, 2) high performance computing utilizing a multi-fidelity hierarchy of physics models ranging from high degree of freedom to reduced models and neural networks, with the aim of both supporting experimental studies and elucidating fundamental physics and 3) validated predictive integrated modeling. Currently, complementary approaches are being pursued in order to achieve the mission of our program (understand, predict, explore, and control):
- **Standalone models**: Individual models ranging from fluid models to full kinetic models can be used to simulate the entire device. Fluid MHD models in particular, such as extended MHD, are beneficial to describe global nonlinear macroscopic plasma behavior, and address the challenges of controlling ELMs and disruptions. Standalone models can also have some loose coupling with external systems. To gain confidence for prediction for reactor scales, high fidelity standalone models should be validated on smaller-scale devices (or simpler experiments). Historically *different MFE magnetic configurations*, including reversed field pinches, spheromaks, and field reversed configurations have been successfully used as validation targets for validation of nonlinear extended MHD (and hybrid kinetic-MHD models). Even non-MFE devices (e.g. LAPD) can play a valuable role in validation at relevant physics parameters and allow further extrapolation.

- **Integrated modeling for fast prediction**: This approach is integrated modeling through direct multiphysics-multiscale coupling of individual high-fidelity models. An integrated model should contain the core confined burning plasma, plasma edge (including scrape-off layer) and the external systems (i.e. plasma facing material, vessel wall, RF antennas, beams, coil controllers).

    There are different types of coupling of high-fidelity codes such as RF-MHD, kinetic core-edge, MHD-kinetic, and edge plasma – multimaterial (coupling EM gyrokinetic to comprehensive models of neutral particle and radiation transport), some of which have been supported by the SciDAC program. These couplings are challenging and require an extensive applied math and computer science effort, which are on the 15-year time line [4]. Reduced-fidelity models calibrated by highest fidelity physics simulations, and experimentally validated, can be used favorably for fast prediction of plasma performance.

By utilizing these approaches, some of *the main objectives of our current program* are 1) to understand and predict the operational limits of the existing experiments, 2) code verification/validation with uncertainty quantification, and 3) physics-based predictive modeling leading to performance optimization and controlling the transients in ITER.

## 4. **Programmatic and Global Context**:

**Partnership with other agencies**: Partnerships with other US agencies, such as ASCR, NSF, NASA enhance the scientific and the educational breadth of the MFE program. Our fusion theory program is very much strengthened by synergy with other subfields of plasma physics as well as other disciplines, such as applied mathematics and computer science.

**International partnership**: The US theory program due to its current scientific leadership in areas such as stellarator optimization and integrated simulation, could further benefit by actively engaging in the rising international fusion programs and the newly built or upgraded fusion experiments around the world: W7X, JET, WEST, JT60-SA, EAST and KSTAR, to name a few. Engaging with the ITER modeling group through the ITPA and the EU, Japanese and Chinese theory communities enhances the productivity of the US MFE theory and computation program.

## 5. Timeline of the Initiative :
## 5.1 Possible 15-year U.S. research agenda

To get the full benefits of the fusion theory program outlined in section 2, we envision that our effort in high performance computing combined with integrated modeling should continue to be pursued for the next 15-years. To achieve theoretical physics-based predictions for a fusion reactor with quantified confidence, at every level, the individual models should be validated on various MFE smaller-scale highly diagnosed devices performing specific validation experiments. There are major challenges and opportunities, which have been identified by the community, as well as critical gaps in theory and numerical methods (and associated gaps in simulation capability) [3]. Here, major opportunities that would have high impact for improving theory and simulation capabilities are:

- **Understand, predict, and control plasma transients:** At present, the number one challenge for burning plasmas, including ITER, is the possibility of disruptions. Advanced simulations need to model all forms of disruption from the initial instability to final wall deposition. Models should be developed to understand and address some of the challenges, including runaway electron generation and evolution, rotation physics and mode locking, disruption-related plasma-wall interaction and open-field currents. Understanding control and mitigation techniques such as, Edge Localized Mode control with external coils, pellet fueling and disruption mitigation, and active control of MHD instabilities are essential. Real-time disruption forecasting from theory-based stability boundary maps and plasma control systems based on neural network and machine learning techniques can potentially provide robust disruption avoidance. In the next 15-years, computational modeling of transients would have a direct impact on the ITER research program.
- **Modeling for long pulse operation:** High performance computation of non-inductive current-drive techniques should be integrated from the edge to the core, showing that current and heat could be deposited in the plasma core and form a steady state. Challenges are predictions for solenoid-free current-drive techniques (through various helicity injection techniques and subsequent RF- neutral beam ramp-up), prediction and mitigation of RF interactions with the plasma-material-interface at the plasma boundary and integrated modeling to predict effective alpha particle heating and possible energetic particle instabilities.
- **Design optimization toward disruption-free configurations:** Optimizations through modeling will guide us to configurations of ultra-optimized stellarators that are inherently steady-state and avoid disruptions. Development of computational tools should be pursued to further exploit the potential of stellarators and to determine the effect of the magnetic configuration on turbulent transport, magnetic surface fragility, macroscopic instabilities, energetic ion confinement, impurity control and edge/divertor physics. Theory and modeling could also investigate 1) the advantages of high temperature superconductors on confinement, and 2) engineering design improvements for advanced divertor, blanket, RF launchers, and outside fluid loops.
- **Improved modeling for plasma-material interaction:** Boundary models should advance to integrate multiple physical processes that cover a wide range of overlapping spatial and temporal scales. This includes integration from the hot, confined pedestal zone with sharp gradients, to the cooler unconfined scrape off layer and divertor plasma where heat fluxes reaching the walls must be within material limits, and finally the first few microns of the wall itself.
- **Understanding of plasma turbulence:** Past and present research into the fundamental processes of instability, saturation, and parametric dependencies of turbulence, using both analytical and numerical tools, will need to be continued. Findings will directly benefit the above categories, help lay the foundation for the next steps beyond the 15-year horizon, and help resolve or mitigate potential problems that may arise for any specific path to fusion energy.

The ultimate goal is to achieve optimization/prediction/control for burning plasmas through whole device modeling (WDM). [3,4] It should be noted that the state of readiness varies for the different elements in the areas outlined above, and some areas would require further resources in order to be ready for integration into a WDM. While a comprehensive assessment of the readiness of different components for a WDM is being performed, continued development of analytical theory combined with validated standalone simulations is still necessary. A whole device model could be an assembly of physics models with a range of fidelity, which ultimately allows simulating from the plasma core to the wall during plasma discharge and from start up to ramp down.  In order to achieve the 15-year physics objectives, advances in mathematical and computational technologies are essential. With the move to exascale computing, further interaction among computer scientists, applied mathematicians and plasma physicists is essential and could ultimately help to overcome the challenges of integrated modeling [3,4].  In particular, predictive modeling could be critical for enabling innovative concepts. A summary of the theory and computation challenges and the required R&D discussed during the Madison and Austin community workshops [5-9] is organized in table 1.

## 5.2   Research directions beyond the 15-year horizon

We envision predictive models of the whole device, which include all components that describe the plasma, from macroscopic equilibrium to micro-turbulence and plasma-surface interactions, and ultimately encompass all components that describe the evolution of a plasma discharge from start-up to termination. Whole device models are required for assessments of reactor performance in order to minimize risk and qualify operating scenarios for next-step burning plasma experiments, as well as time-dependent or single-time-slice interpretive analysis of experimental discharges. In ITER, the exploration of burning plasma and pulse control scenarios will be guided by modeling, as it is not feasible to determine operational limits by running trial discharges. The goals of the theory program should include models that provide 1) options for interpretive as well as predictive modes, and synthetic diagnostics, 2) an environment for connecting to experimental databases, possibly on remote platforms, and 3) infrastructure to support the above, as well as for machine and scenario design and operation.

| Objective | Challenges | R&D needed/Gaps |
|---|---|---|
| **Understand, predict, and control plasma transients** | • Disruptions and runaway electron generation and evolution, rotation physics and mode locking, disruption-related plasma-wall interaction and open-field currents<br>• Pellet fueling and disruption mitigation<br>• Power threshold for the H-mode transition<br>• Edge Localized Mode control with external coils<br>• Active control of MHD instabilities | • Further development of analytical theory and validated modeling for all challenges listed in the left column<br><br>Innovations  in the areas of  integrated coupled models:<br>❖ High-fidelity coupling of core-pedestal-SOL system through kinetic (EM gyrokinetic – full kinetic) or MHD-kinetic core-edge coupling for transients such as ELM growth and ejection, and stabilizing physics effects of energetic particles and runaway electrons<br>❖ Real-time disruption forecasting from theory-based stability boundary maps<br>❖ Plasma control systems based on neural network and machine |
| | | learning techniques to provide robust disruption avoidance |
| **Modeling for long pulse operation (Scenarios)** | <ul><li>Steady-state coupling of core, edge, and plasma material interactions</li><li>Fast ion instabilities and transport</li><li>Interaction of fast particles with thermal plasma waves</li><li>Explore operation of DT fusion devices with aneutronic fuel mixtures (for example by replacing the tritium with helium-3)</li></ul> | <ul><li>Validated predictive extended MHD simulations for non-inductive "solenoid-free" current-drive (through various helicity injection techniques and subsequent RF- neutral beam ramp-up)</li><li>Modeling to investigate high-field LHCD launch and its impact on microturbulence in the SOL</li></ul><br>Innovations in the areas of integrated coupled models:<ul><li>A predictive capability for self-consistent interaction of RF power with the scrape-off layer and wall, including realistic antenna and first wall geometry</li><li>Integrated modeling to predict effective alpha particle heating and possible energetic particle instabilities</li><li>Investigate additional heating requirements for aneutronic fusion</li></ul> |

| **Design optimization toward disruption-free configurations** | <ul><li>Stellarator fast-ion and thermal confinement optimization</li><li>Impact of high-Tc superconducting magnets on confinement configurations</li><li>Explore new magnetic configurations</li></ul> | <ul><li>Further development of analytical theory and simulations for:</li></ul><ul><li>improved stellarator optimization</li><li>evaluating the implications of HTS on stability and the heat flux width</li><li>compact tokamak/ST design to lower aspect ratio for greater magnetic field utilization, improved stability, and reduced TF magnet mass</li><li>optimization of all existing MFE concepts to assess their potential for improved stability and confinement and to explore new magnetic concepts</li></ul><p>Innovations in the areas of integrated coupled models:</p><ul><li>Develop computational tools to couple EM GK codes to 3-D (MHD) equilibrium conditions for the purpose of minimizing turbulence in stellarators</li><li>Development of nonlinear MHD and further development of transport codes (such as TASK3D) for stellarators</li><li>Integrated physics and engineering optimization design for advanced divertor, blanket, RF launchers, and outside fluid loops for reactor design and safety</li></ul> |
|---|---|---|
| **Improved modeling for plasma-material interaction** | <ul><li>Reliably predict scrape-off layer transport and beyond</li><li>Plasma material interaction</li><li>Material resilience to neutron damage</li></ul> | <ul><li>Develop codes to examine advanced divertor concepts, including alternate magnetic-geometry divertors and liquid walls</li></ul><p>Innovations in the areas of integrated coupled models:</p><ul><li>Multi-scale SOL models including molecular dynamics and kinetic Monte Carlo codes, 2D and 3D plasma transport codes, and 4-5D EM</li></ul> |

|  |  | <ul><li>gyrokinetic codes</li><li>Plasma codes to couple with efficient wall models for erosion/redeposition of surfaces, impurity release, and tritium trapping within the wall</li><li>Liquid walls.</li></ul> |
|---|---|---|

**Table 1: Summary of the theory and computation challenges in the four areas in Section 5.**

## 6. Cost Range

While the US fusion theory program is considered very successful, past decades have seen some inflation-adjusted budget tightening. In order to fulfill the initiative goals and ensure a successful overall fusion program, these restrictions need to be lifted. In addition, new components such as real-time control, turbulence optimization, multi-scale SOL models and wall solutions, as well as areas listed in Table 1, with needed innovations and R&D, will require additional personnel. The theory and computation budget should therefore be increased in proportion to the needs of the future fusion energy program. A 50% increase in the Base US Theory and Computation Budget would significantly enhance our ability to realize the goals of this initiative. More specifically, this would allow domain (fusion) scientists to address critical gaps in theory and to work with applied mathematicians and computer scientists on the co-design of a new generation of validated and predictive model hierarchies needed for a Whole Device Model; thus facilitating routine analysis of experiment, scenario-development, and shot-planning needs, as well as providing predictive capability for the design of next-step machines.

## 7. Cross-Cutting Opportunities

Many connections to other efforts have been alluded to throughout this initiative. The opportunities listed in section 5 present some of the theory and computational key gaps in the main areas of transients, energetic particles, transport and confinement, scenarios, and plasma and material interaction. By virtue of both supporting the role of theory for experiment as well as the enabling properties of fundamental physics understanding in overcoming newly-arisen obstacles, a healthy theory program will benefit all pathways to fusion energy. More specifically, this includes but is not limited to sustained high-power tokamak initiatives (Bonoli, Menard, Buttery, Kessel, Greenwald, Battaglia, Bongard), optimized stellarator initiatives (Gates, Bader, Zarnstorff, Lazerson), alternate configurations (Sutherland, McCollam, Prater), and the negative-triangularity configuration space (Austin). Select components of the present initiative share common space with those proposed in the Meneghini initiative, and exascale and capacity computing (Bhattacharjee and Lyons) initiatives. Our initiative aims to stress the key role of a range of complementary tools from analytical theory to high performance computing utilizing a multi-fidelity hierarchy of physics models, which could help to close the identified gaps in order to furtherance toward fusion energy.

**Referees:** David Newman (U. Alaska) and John Canik (ORNL)

**Advocates:** B. Grierson (PPPL) and W. Horton (UT), D. Brennan (Princeton University), Alex Friedman (LLNL)

**Revision History**

- Version 3: September 27 2019 - Revised Initiative for Knoxville APS-DPP-CPP workshop
- Version 2: September 2019 - Per request from members of theory & computations CC group resubmitted version 1 to APS-DPP-CPP
- Version 1: 29 February 2018 - Submitted reviewed (by Newman and Canik) whitepaper as one of the main community-led whitepapers submitted to NAS on strategic elements